\documentclass[aps,twocolumn,nofootinbib]{revtex4}

\usepackage{graphicx}
\usepackage{amssymb, amsmath,mathtools}
\usepackage{bbm}
\usepackage{enumerate}
\usepackage{xcolor}
\usepackage{slashed} 

\graphicspath{{./}{./plots/}}

\usepackage{xcolor}

\begin{document}

\title{Majorana quasiparticles, mass hierarchies, and the phase diagram of the Gross-Neveu model in D=2+1}

\author{Igor F. Herbut and Mathew Schneider}

\affiliation{Department of Physics, Simon Fraser University, Burnaby, British Columbia, Canada V5A 1S6}

\begin{abstract} 

We show that the canonical $2+1$-dimensional $O(2N)$-symmetric Gross-Neveu model with the inversion-breaking mass-term features a sequence of $O(2N-k) \times O(k)$ - symmetric ground states in its zero-temperature phase diagram, with $k=1,...,N$. For odd $k$ the quasiparticle excitations are itinerant, massive, two-real-component Majorana fermions, i.e. half of Dirac fermions. The solutions of the mean-field equations symmetric under $O(2N-k_1-k_2) \times O(k_1) \times O(k_2)$ with $k_1, k_2 \geq 1$ also exist, but are found to be unstable. Some manifestations of Majorana quasiparticle  in the particular ``uniaxial nematic" state with $k=1$ are outlined, and analogies to other physical systems are offered.
\end{abstract}

\maketitle

{\it Introduction} - The Gross-Neveu (GN) field theory is maybe the simplest model of interacting relativistic fermions, and as such has played a crucial role in both high-energy and condensed matter physics. It was initially proposed as a toy model in $D=1+1$ dimensions that exhibits asymptotic freedom and dynamical fermion mass-generation.  \cite{GrossNeveu} Its various reduced-symmetry modifications in $ D=2+1$ are considered today to be low-energy descriptions near quantum phase transitions of several realistic systems of interacting electrons, such as monolayer graphene, twisted bilayer graphene, or d-wave superconductors. \cite{HerbutReview} The original $D= 2+1$ GN model for $N$ two-component Dirac fermions, at weak coupling, besides the usual space-time symmetries such as Lorentz, inversion, and time-reversal, also has the internal orthogonal symmetry $O(2N)$. \cite{HerbutMandal} As the magnitude of the single coupling constant of certain sign, which we define here as negative, in the GN theory is increased, at a critical value $\sim 1/N$, the inversion and the time-reversal symmetries spontaneously break, and the spectrum becomes massive. At this phase transition the $O(2N)$ symmetry is preserved, which facilitates the large-N expansion as the analytic method of choice. \cite{ZinnJustin} All two-component fermions acquire the same mass in the broken symmetry phase.  In graphene terminology, for example, this mass would be the order parameter for the quantum anomalous Hall phase.\cite{HuangLee}  The quantum critical point for the inversion- and time-reversal symmetry breaking transition has been studied by large-N, \cite{Rosenstein, Vasilev, Gracey1, Gracey2}, Monte Carlo \cite{Karkkainen},  the expansions near the lower and upper critical dimensions, \cite{Schroeder, ZinnJustinNPB, Fei, Zerf}, exact renormalization group methods \cite{Rosa, Hofling},  and more recently by conformal bootstrap. \cite{Erramilli}  It represents probably the best understood fermionic criticality to date.

A weak {\it positive} coupling in the GN theory is, on the other hand, found in perturbation theory to be irrelevant in the infrared. Nevertheless, at higher positive coupling one expects another phase transition at which, besides parity, the internal $O(2N)$ symmetry breaks as well. This expectation is based on the  exact transformation of the original GN interaction term of the form of scalar$^2$  in Eq. (3) into the form of tensor$^2$ (Eq.(4)),  and the concomitant mean-field calculation. \cite{HerbutMandal} Further analysis near the upper critical dimension of $D=3+1$, \cite{Han1, Han2} together with the Monte Carlo study of the appropriate lattice model \cite{Rein} suggest the transition at the critical positive coupling likely to be fluctuation-induced first-order. The $O(2N)$ symmetry at this transition is broken down to $O(N) \times O(N)$, and consequently all fermionic excitations end up again having mass of the same magnitude, only this time with alternating signs. They all can therefore be combined into $N/2$ four-component massive Dirac fermions. In the case of graphene, this phase may be understood as the usual charge density wave (CDW), for example. \cite{HerbutReview, HerbutPRL, HerbutJuricicRoy}

\begin{figure}[t]
{\centering\resizebox*{70mm}{!}{\includegraphics{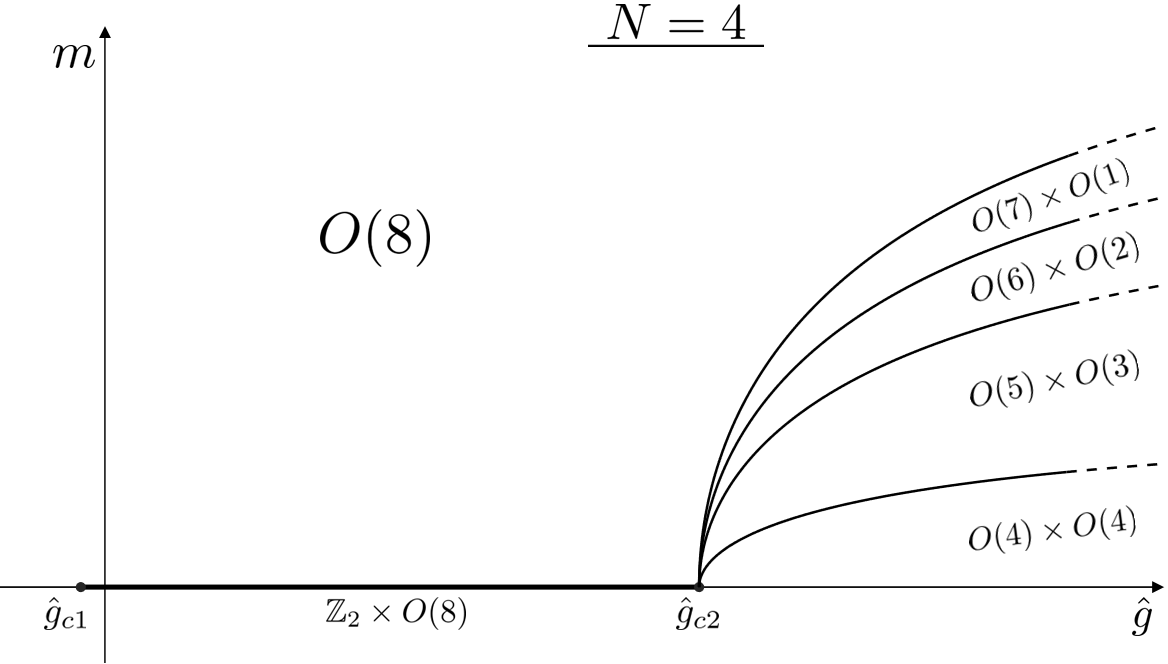}}
\par} \caption[] {Phase diagram of the GN model (Eqs. (2) and (3)) for $N=4$. For $\hat{g}_{c1} < g < \hat{g}_{c2}$ and $m=0$ the model has the inversion ($Z_2$) and the time-reversal  symmetry, whereas everywhere else the inversion and time reversal are broken, spontaneously (at $m=0$), or explicitly (at $m\neq 0$). All the phase transitions are discontinuous, except at $\hat{g} = \hat{g}_{c1}$, $m=0$, and possibly at $\hat{g} = \hat{g}_{c2}$, $m=0$.}
\end{figure}

When the GN model describes a lattice system of fermions, the above flavor number $N$ is typically an even integer, as a consequence of the ubiqutous  fermion-doubling.  \cite{Nielsen, Ninomiya, Janssen} The two massive broken-symmetry phases mentioned above both feature then the usual Dirac fermions as the quasiparticle excitations, with equal or sign-alternating masses, depending on whether $O(2N)$ symmetry is broken, for positive, or not, for negative interaction. One may wonder if the outcome could possibly be any different, and if more exotic possibilities for the quasiparticle excitations may be realized. In this Letter we provide a surprising and affirmative answer to this question. Starting from the $O(N) \times O(N)$ symmetric phase above the positive critical coupling, we find that by increasing the parity-breaking $O(2N)$-symmetric mass-term, the system  goes through a whole series of other broken-symmetry phases on its way to full restoration of $O(2N)$ (see Fig. 1): 
\begin{widetext}
\begin{equation}
  O(N) \times O(N) \rightarrow O(N+1) \times O(N-1)\rightarrow ... O(2N-1)\times O(1) \rightarrow O(2N). 
\end{equation}
\end{widetext}
In the ground states with $O(2N-k)\times O(k)$ symmetry the excitations are $k$  and $2N-k$ massive {\it real} (Majorana) two-component fermions with masses of two different magnitudes. In particular, when $k$ is odd, combining them into a usual Dirac fermion is not possible. The transition from the $O(2N)$ symmetric phase is always into a phase that has only a {\it single} such massive Majorana fermion. The fermionic excitation spectrum in this phase is therefore similar to the one in Kitaev model in magnetic field, although the physical origin of the Majorana fermions is entirely different. \cite{Kitaev} 

{\it GN model} - The GN model in $D=2+1$ is defined by the zero-temperature ($T=0$) action $S=\int d\tau d^2 \vec{x} (L_0 + L_{int})$,
\begin{equation}
L_0=  \phi^{\top}  1_{2N} \otimes (1_2 \partial_\tau -i \sigma_1 \partial_1 - i \sigma_3 \partial_2  + m\sigma_2 )   \phi ,
\end{equation}
\begin{equation}
L_{int} = g (\phi^{\top} 1_{2N} \otimes \sigma_2 \phi)^2,
\end{equation}
where $\tau$ is the imaginary time, and $1_M$ is the $M$-dimensional unit matrix. The cutoff $\Lambda$ on the momentum  and frequency integration is assumed. In our units $\hbar = k_B = c =1$, where $c$ is the velocity of fermions. $\phi=\phi(\vec{x}, \tau) $ 
is the $4N$-component real fermion $\phi^{\top} = (\phi_1 ^{\top}, \phi_2 ^{\top} )$. The non-covariant, ``real"  representation above makes the internal symmetry $\phi \rightarrow (O \otimes 1_2) \phi$ with $O \in O(2N)$ explicit.   The usual complex fermion may also be defined as 
 $\psi = (\phi_1 - i \phi_2)$ and $\psi^\dagger = (\phi_1 ^{\top}+ i \phi_2 ^{\top})$, to rewrite the GN model in the standard covariant form, 
 $L= \bar{\psi} (\slashed{\partial} +m) \psi + g (\bar{\psi} \psi)^2 $, with $\bar{\psi}=\psi ^\dagger 1_N \otimes \sigma_2$. Besides Lorentz symmetry, at $m=0$ the model is symmetric under inversion $\psi \rightarrow 1_N \otimes \sigma_1 \psi$, $(x_1,x_2,\tau)\rightarrow (x_1,-x_2,\tau)$, and time-reversal  $\psi \rightarrow 1_N \otimes \sigma_2 K \psi $, with $K$ as complex conjugation. 
 
 When $m=0$, for the dimensionless coupling $\hat{g} = g \Lambda <\hat{g}_{c1} \sim -1/N$ (at large $N$) the inversion and the time-reversal symmetries are spontaneously broken by the condensation of the $O(2N)$-singlet: $\langle \phi^{\top} 1_{2N} \otimes \sigma_2 \phi \rangle  \neq 0$. When $m=0$ and $\hat{g}> \hat{g}_{c1}$, in contrast, $\langle \phi^{\top} 1_{2N} \otimes \sigma_2 \phi \rangle = 0$.  $\langle \phi^{\top} 1_{2N} \otimes \sigma_2 \phi \rangle  \neq 0$ at all $m\neq 0$, when the inversion  and the time-reversal symmetries are manifestly broken. 
 
 It has been previously established \cite{HerbutMandal, HanHerbut} that the interaction term can be rewritten exactly as:
 \begin{equation}
 L_{int} = -\frac{g}{N+1} (\phi^{\top} S_b \otimes \sigma_2 \phi) (\phi^{\top} S_b \otimes \sigma_2 \phi).
 \end{equation}
 The summation over the repeated indices is hereafter assumed, and the index $b=1,2,...(N+1) (2 N-1)$ enumerates linearly independent $2N \times 2N$, traceless, symmetric real matrices $S_b$, normalized as  $Tr (S_a S_b) = 2 N \delta_{ab}$. The negative sign appearing above suggests that for  $\hat{g} > \hat{g}_{c2} \sim 1 $ and $m=0$  $O(2N)$ spontaneously breaks down to $O(N) \times O(N)$, and some components of the symmetric tensor order parameter condense,  $\langle \phi^{\top} S_b \otimes \sigma_2 \phi \rangle \neq 0$.\cite{HerbutMandal, Han1, Rein, HanHerbut}

{\it Phase diagram} - We are primarily interested in the ground states of the GN field theory when $g > 0$, and $m>0$. After decoupling of $L_{int}$ in Eq. (4) with Hubbard-Stratonovich real fields $\chi_b$, the integration over fermions yields the compact expression in the continuum limit: 
\begin{equation} 
\frac{12 \pi}{m^3} L(S) = a Tr S^2 + Tr | 1_{2N} + S|^3.
\end{equation}
Here we assumed a constant (dimensionless) matrix $S= \chi_b S_b /m$, omitted the $S$-independent terms, and took the limit $\Lambda \rightarrow \infty $. (see End Matter.) The dimensionless coefficient of the quadratic term is 
\begin{equation}
a= \frac{3 \Lambda}{\pi m} ( \frac{ \pi^2 (N+1) }{2 N \hat{g} } -1).
\end{equation}

When $m\rightarrow 0$, $L(S)$ has a non-trivial stable minimum at $S= -(2 a /3) \sigma_3 \otimes 1_N$ for $ \hat{g}  > \hat{ g} _{c2} = (\pi^2/2 ) ( 1 + (1/N) )$, at which $O(2N) \rightarrow O(N) \times O(N)$. \cite{HerbutMandal} 

\begin{figure}[t]
{\centering\resizebox*{70mm}{!}{\includegraphics{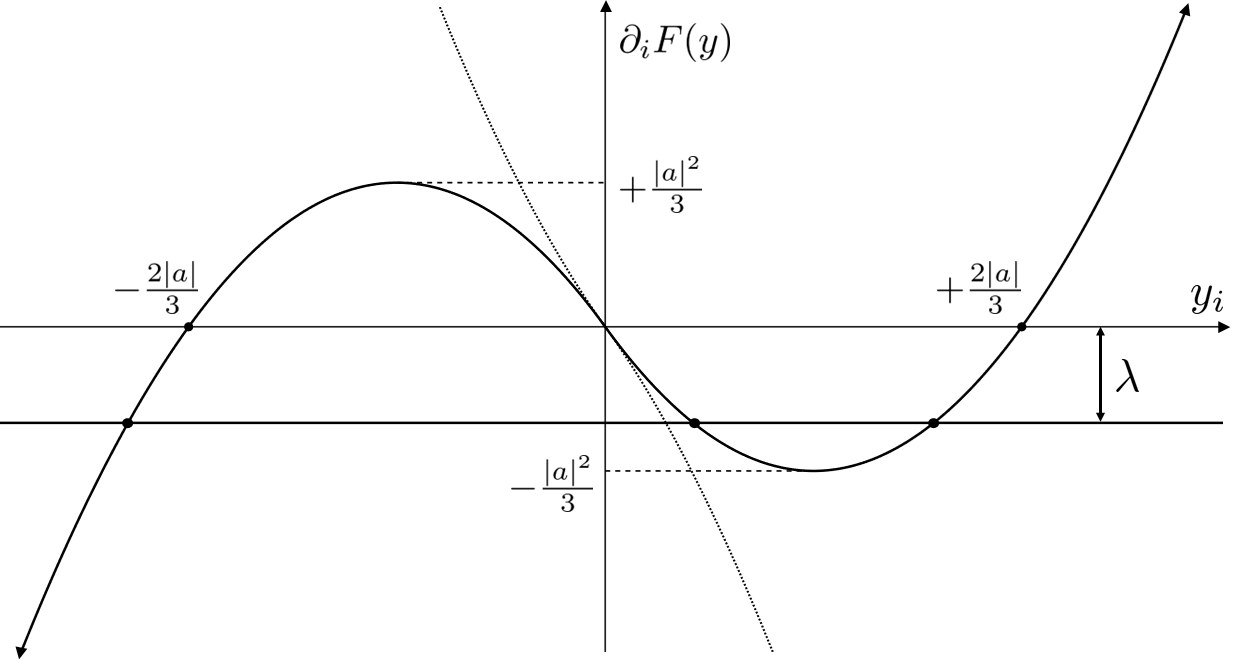}}
\par} \caption[] { The solution of Eqs. (8) for $a<0$ and $\lambda < 0$ lies at the intersection of the horizontal line and the full-line sections of the two parabolas. }
\end{figure}

The minimization of $L(S)$ at finite $m$ is more involved, and there are many local extrema. Call $x_i$, $i=1,...2N$, the eigenvalues of the matrix $S$; then $Tr S= \sum x_i =0$. The function to be minimized then becomes:  
\begin{equation}
F(y) = \sum_{i=1} ^{2N} (a  y_i ^2 +  |y_i|^3),  
\end{equation}
with $y_i =1+x_i$. The equations for the extrema are: 
\begin{equation}
2 a y_i + 3 y_i |y_i| =\lambda, 
\end{equation}
with $\lambda$ as the Lagrange multiplier. 

Graphical solution of these equations is depicted in Fig. 2, for $a<0$. There can evidently be at most three distinct values of $y_i$.  For $a>0$ the only solution is $y_i \equiv 1$, i.e. the trivial matrix $S=0$. (End Matter.) We find that the local minima of $L(S)$ at $a<0$ are always at $S$ with only {\it two} distinct eigenvalues: $y_i =u>0$ with degeneracy $2N-k$, and $y_i =v <0$ with degeneracy $k$. The extrema with three distinct values of $y_i$, when they exist, are always saddles. (End Matter) In the case of two distinct $y_i$, combining the extrema equations with the constraint  
\begin{equation}
(2N-k) u + k v = 2N,
\end{equation}
one solves explicitly for $u$ (and the corresponding $v$): 
\begin{widetext}
\begin{equation}
u= \frac{2}{ 3 (k^2 + (2N-k)^2) } [ - N ( a k -3(2N-k)) + \sqrt{ N^2 ( a k -3 (2N-k))^2 - 3 N  (k^2 + (2N-k)^2) (3 N - a k ) } ].  
\end{equation}
\end{widetext}
The solutions are real only for 
\begin{equation}
a< a_{sp} (k) = \frac{3}{N} ( (N-k) - \sqrt{ (N-k)^2 + N^2 } ). 
\end{equation}
The ``spinodal" point value $a_{sp} (k)$ is evidently a negative, decreasing function of $k$. Decreasing the coefficient $a$ from zero toward negative values the $k=1$ solution becomes real first at $a= a_{sp} (1)  $, and then becomes the first non-trivial global minimum at $a=a_c (1) < a_{sp} (1) $. At a finite $m$ at the transition from the symmetric phase the symmetry is therefore broken as  
\begin{equation}
O(2N) \rightarrow O(2N-1) \times O(1), 
\end{equation}
with the $O(1)$ to be understood as $Z_2$.

For the maximal value of $k= N$, $a_{sp} (N) = -3$, independent of $N$; then $u=2$, $v=0$, and $\lambda=0$. 

\begin{figure}[t]
{\centering\resizebox*{70mm}{!}{\includegraphics{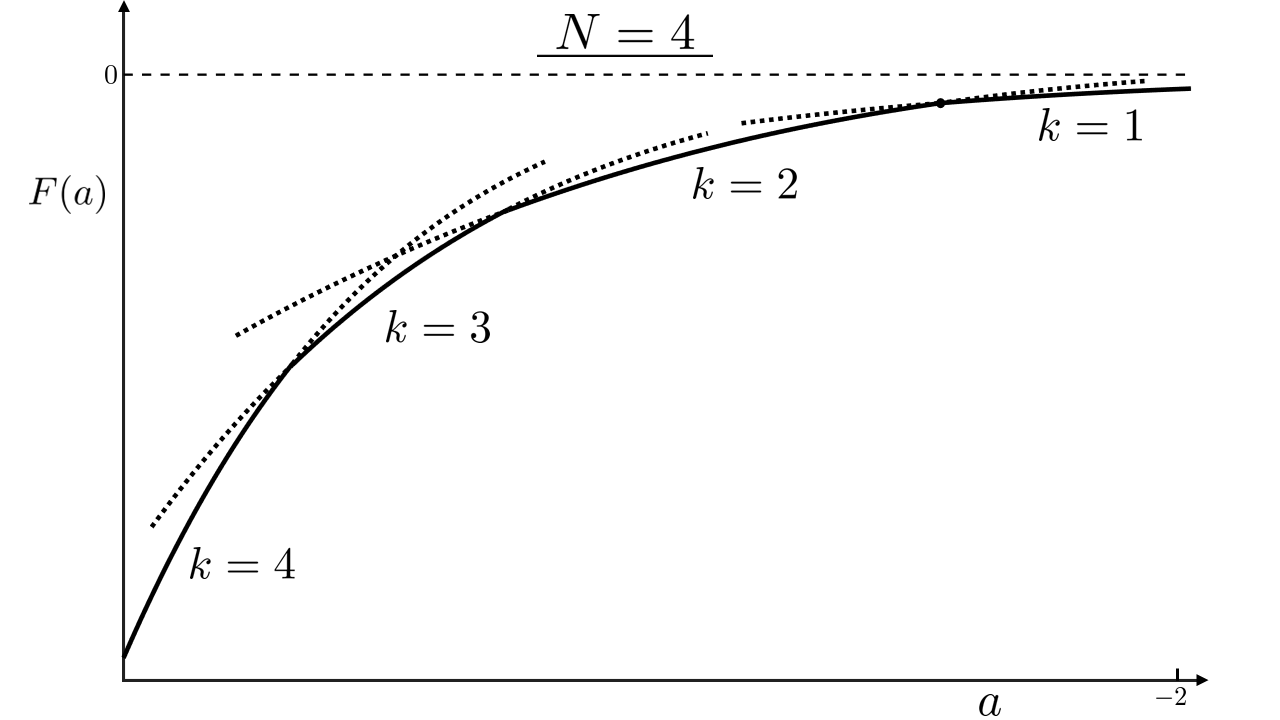}}
\par} \caption[] { Functions $F(a)$ for different $k$ and $N=4$. The ground state is determined by the lowest laying curve in the given interval of $a$. The crossings accur at $a= -1.72, -2.4, -4.02, -12$. }  
\end{figure}

There are therefore at least two different broken symmetry ground states for all $N\geq 2$ at $g>g_{c2}$:  the $O(N) \times O(N)$-symmetric phase for $m=0$ and $m$ infinitesimal, and the $O(2N-1) \times O(1) $-symmetric phase for $m$ infinitesimally below  the line $(m/\Lambda= 3(g-g_{c2})/(\pi g | a_{c}(1) |)$.  When $N=2$, there are no other broken symmetry ground states at $g>0$. For higher $N$, however, two-eigenvalue solutions with other degeneracies eventually become competitive in energy. Inserting the solutions for $u$ and $v$ for different $k$ back into $F(y)$, we obtain the set of intersecting curves as on Fig. 3. With an increase of $m$ at fixed $\hat{g} > \hat{g}_{c2}$, for any $N$ there is a series of level crossings with the symmetry changing as announced in Eq. (1). The resulting $g-m$ phase diagram of the GN model is depicted on Fig. 1. 

{\it Uniaxial nematic} - The direct transition from the $O(2N)$- symmetric state is thus always into the $O(2N-1) \times Z_2$ - symmetric state, which in analogy with the state of liquid crystals \cite{deGennes} we name {\it uniaxial nematic}. The symmetry breaking pattern implies  $2N-1$ gapless Goldstone bosons. The resulting mass-matrix $S_{nem}$ is a sum of $2N-1$ {\it commuting} matrices $S_b$ with equal coefficients. Besides the Goldstone bosons at low energy, the excitations are one two-component Majorana fermion with the mass $v m$, and $2N-1$ two-component Majorana fermions with the mass $u m$. Near the transition we find  $|v|< u$, and therefore one ``light" and $2N-1$ ``heavy" Majoranas. The  leading contribution of the light Majorana fermion to low-temperature  specific heat is, for example, 
\begin{equation}
E_M =\frac{1}{2} \frac{|v|^3 m^3}{\pi T } e^{-\frac{ |v|m}{T} }, 
\end{equation}
which is {\it a half} of the usual two-component Dirac result. The contribution from the heavy Majoranas is exponentially suppressed, and the Goldstone bosons yield $\sim T^2$.   

To illustrate further the uniaxial nematic ground state we consider the simplest case $N=2$. The $4 \times 4$ real matrix $S$ can then be conveniently written as 
 \begin{equation}
 S= c_{ij} M_{ij},
 \end{equation}
 with $(M_{11},M_{12},M_{13}) = (1\otimes\sigma_1, 1\otimes\sigma_3, \sigma_2 \otimes\sigma_2)$, $(M_{21},M_{22},M_{23}) = (\sigma_1\otimes\sigma_3, \sigma_1\otimes\sigma_1, \sigma_3 \otimes 1)$, and $(M_{31},M_{32},M_{33}) = (\sigma_3\otimes\sigma_3, \sigma_3\otimes\sigma_1, \sigma_1 \otimes 1)$, and $c_{ij}$ real. The nine matrix-components of $S$ form a symmetric irreducible two-component tensor under $O(4) \simeq SU_L (2) \times SU_R (2)\times Z_2$, with the $SU_L(2)$ generated by  $L_i= (1\otimes\sigma_2, \sigma_2\otimes\sigma_3, \sigma_2 \otimes \sigma_1)/2$ and the $SU_R (2)$ by $R_i= -(\sigma_3\otimes \sigma_2, \sigma_2\otimes 1, \sigma_1 \otimes \sigma_2)/2$. The matrices $M_{ij}$ are defined so to anticommute only if they share the same value of an index. A matrix $S_{nem}$ is proportional to a sum with equal coefficients of any three $M_{ij}$ that do not share an index and therefore commute; for example, $S_{nem} = \chi ( M_{12}+M_{21}+M_{33})$.  General $3\times 3$ matrix of coefficients $c_{ij}$ in Eq. (14) in the uniaxial nematic state is proportional to an {\it orthogonal matrix}.  Although $S_{nem}$ breaks all of the individual generators $L_i$ and $R_i$, it preserves a subgroup  $SU_+ (2)$ generated by  $O^T (L_i + R_i) O$, $i=1,2,3$, where the orthogonal matrix $O$ diagonalizes the given $S_{nem}$; $S_{nem} = \chi O^T D  O$,  with $D$ diagonal: $D_{11}=3$, $D_{22}=D_{33}=D_{44} = -1$ and $D_{i\neq j} =0$.  The pattern of symmetry breaking for $N=2$ is reminiscent of that of the Balian-Werthamer superfluid state of $^3 He$, \cite{Wolfle} where independent rotations in the spin and orbital Hilbert spaces in the normal phase are reduced to the joint $SO(3)$ that rotates both degrees of freedom together.  
 
 {\it Discussion} - We have considered the standard GN model in $D=2+1$ dimensions in presence of the inversion and time-reversal symmetry violating and flavor-symmetry respecting mass $m$ and at $T=0$. Taking the sign and magnitude of the interaction that favor the spontaneous breaking of the flavor $O(2N)$ symmetry at $m=0$, the mean-field theory at $m\neq 0$ leads to a sequence of ground states with symmetries $O(2N-k) \times O(k)$, $k=1,...N$. The transitions between the states that correspond to different values of $k$ are discontinuous. Most interestingly, the fermionic excitations above these ground states are massive Majorana fermions with masses of two distinct magnitudes. When $k$ is odd these Majorana quasiparticles cannot be combined into usual Dirac fermions, and represent fractionalized excitations.     
 
 It is instructive to contrast the phase diagram found here with what would occur without the $O(2N)$ symmetry, if the sum over all tensor components in Eq. (4) is replaced by a single term, with $S_1 = \sigma_3 \otimes 1_{N} $, for example. In graphene this matrix may be understood as representing density imbalance on two sublattices, i. e. CDW. Increasing the mass $m$ in the state with $\langle \phi^T S_1 \otimes    \sigma_2 \phi \rangle \neq 0$ would then cause a direct phase transition from the (topologically) trivial insulator with $O(N) \times O(N)$ symmetry to the non-trivial $O(2N)$-symmetric insulator \cite{Haldane}  at the value of $a=-3$, identified below Eq. (12). For $N=4$ this value lies within the $O(6) \times O(2) $ phase, for example. (Fig. 3)  In the $O(2N)$-symmetric GN model this direct transition is avoided by the order parameter choosing other available directions in the order parameter space before the symmetry gets completely restored at higher $m$. An (incomplete) analogy may be drawn with the response of the N\' eel order parameter to the magnetic field; the N\' eel vector chooses the direction orthogonal to the magnetic field, if available, to minimize the loss in the correlation energy caused by the  finite magnetization. 
 
Finally, one lesson of the present calculation may be that a hierarchy of the masses may be generated by standard spontaneous symmetry breaking under suitable  conditions. In the GN model in $D=2+1$ not more than three different magnitudes of the masses are possible, and only two are found in the ground states. Nevertheless, we observe that in the minimal theory with $N=2$, for example,  one finds one light and three heavy Majorana fermions near the $O(4) \rightarrow O(3) \times O(1)$ transitions; near the second, $O(3) \times O(1) \rightarrow O(2) \times O(2)$ transition, it is the other way around and the mass hierarchy is ``normal".\cite{zee}   It remains to be seen  if a similar mechanism may be relevant for the problem of the neutrino masses.

 {\it Acknowledgement} - IFH is grateful to Yuan-Ming Lu, Lukas Janssen, Michael Scherer, and  Max Uetrecht for useful discussions. This work was supported by the NSERC of Canada. 
 
 {\it Data availability} - No data was generated or analyzed in this manuscript. No AI was used either.

\begin{widetext}
\section{End Matter}
\end{widetext}

{\it Integration over fermions} -  After the Hubbard-Stratonovich transformation for $g>0$ the Lagrangian $L= L_0 + L_{int} $ becomes
\begin{eqnarray}
L = \frac{N+1}{4 g} \chi_b \chi_b +  \phi^{\top}  [1_{2N} \otimes (1_2 \partial_\tau -i \sigma_1 \partial_1   \\ \nonumber
- i \sigma_3 \partial_2  + m\sigma_2 )  
+ (m 1_{2N} + \chi_b S_b)\otimes \sigma_2 ]  \phi.  
\end{eqnarray}
Assuming the minimum to be at uniform fields $\chi_b$ the fermions can be integrated out, so that 
\begin{equation} 
L = \frac{N+1}{8 N g} Tr S^2 - \frac{1}{2}  Tr \int \frac{d^3 Q }{ (2\pi)^3 }  \ln (Q^2 1 _{2N} + ( m 1_{2N} +S)^2 ),  
\end{equation}
where $S=\chi_b S_b$, $Tr S=0$, $Tr S_a S_b = 2N \delta_{ab}$.  Introducing the eigenvalues $x_i$, $i=1, ... 2N$ of the real symmetric matrix $S$, the Lagrangian becomes 
\begin{equation}
L= \frac{N+1}{8 N g} \sum_i x_i ^2 - \frac{1}{4 \pi^2} \sum _i \int _0 ^\Lambda dQ Q^2 \ln (Q^2 + (m+ x_i)^2 ).
\end{equation}
Partial integration then gives 
\begin{eqnarray}
L= \frac{N+1}{8 N g} \sum_i x_i ^2 - \frac{1}{4 \pi^2} \sum_i [ \frac{\Lambda^3}{3} \ln (\Lambda^2 + (m+x_i)^2)  \\ \nonumber
 -\frac{2 \Lambda^3 }{ 9} + \frac{2 \Lambda (m+x_i) ^2}{3} -\frac{2 (m+x_i) ^4}{ 3 |m+x_i|} \arctan \frac{\Lambda}{|m + x_i|} ].  
\end{eqnarray}
Assuming that $ |m + x_i|/\Lambda \ll 1$ and expanding in powers of small $m+x_i$ yields
\begin{eqnarray}                                                                                                                                                                                                                                                                                                     L= \frac{N+1}{8 N g} \sum_i x_i ^2 - \frac{1}{4 \pi^2} \sum_i [  \frac{\Lambda^3}{3} ( \ln \Lambda^2 - \frac{2}{3} )   \\ \nonumber
+ \Lambda (m+x_i ) ^2 - \frac{\pi}{3} |m+ x_i|^3 +O((m+x_i) ^4 /\Lambda) ]. 
\end{eqnarray}
Since $\sum_i x_i =0$, omission of the terms independent of $x_i$ gives  
\begin{widetext}
\begin{equation}
L= ( \frac{N+1}{8 N g} -\frac{\Lambda}{4 \pi^2} ) \sum_i x_i ^2 + \frac{1}{12\pi} \sum _i |m+ x_i|^3 + O((m+x_i) ^4/\Lambda). 
\end{equation}
\end{widetext}
Finally, at $m\neq 0$, rescaling $x_i$ as $x_i \rightarrow x_i/m$ and taking the continuum limit one arrives at the Eqs. (5) and (6) in the main text. 

{\it Solutions for the extrema} - At $a>0$, assuming the Lagrange multiplier $\lambda>0$ leads only to one positive solution for the eigenvalues, call it $u$. (see Fig. 4) Since all eigenvalues are then the same, the constraint is then  $2N u = 2N $, 
so that $u=1$, which implies that $S\equiv 0$, i. e. trivial. Assuming $\lambda <0$ would lead to all eigenvalues being the same but negative, which cannot satisfy the constraint. So the $O(2N)$ symmetry is preserved for the quadratic coefficient $a>0$. 

\begin{figure}[t]
{\centering\resizebox*{60mm}{!}{\includegraphics{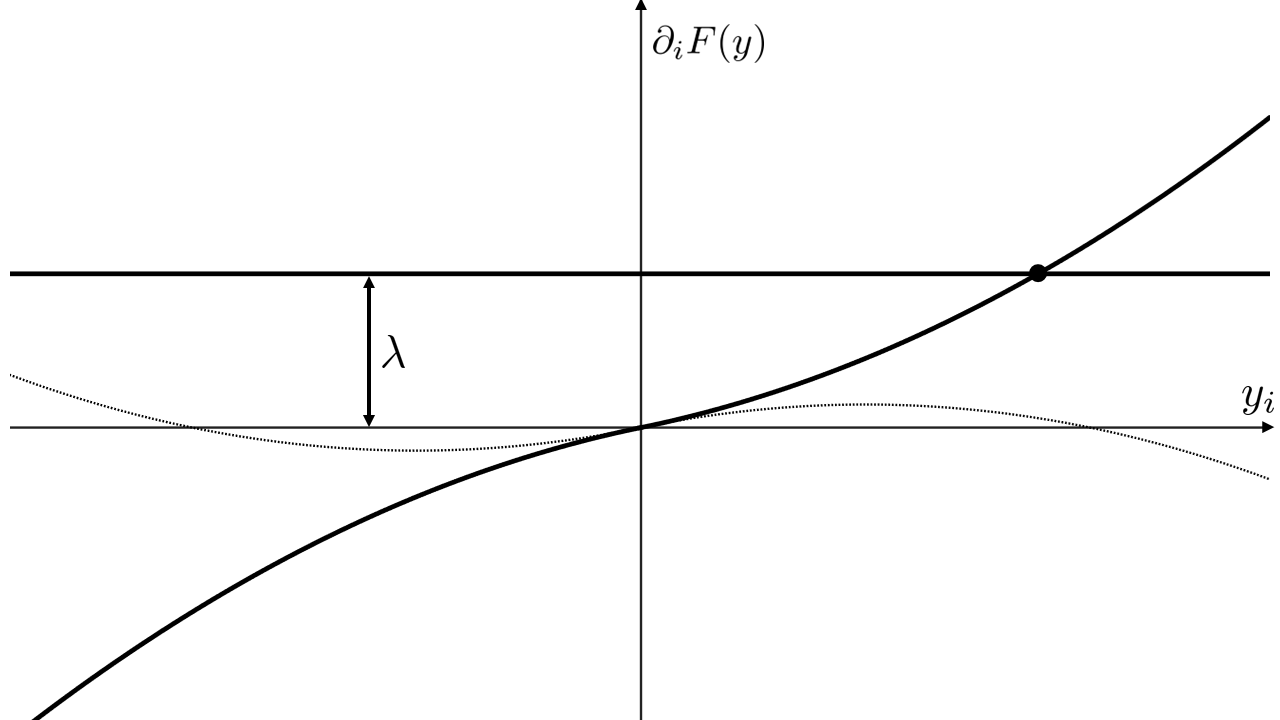}}
\par} \caption[] { Graphical solution of Eqs. (8) for $a>0$ and $\lambda>0$. }

\end{figure}

The case of $a<0$ is treated explicitly in the paper, for the case of two distinct eigenvalues. Let us show that the extrema with three distinct eigenvalues are always saddles. First, let us assume we have $k_1$ eigenvalues $y_i = u >0$, $k_2$ with $y_i =v<0$, and $k_3$ with $y_i=w$, of either sign. Obviously 
$k_1 + k_2 + k_3 = 2N$, and $k_i \geq 1$, $i=1,2,3$. 

From Fig. 2, we see that the highest possible value of Lagrange multiplier $\lambda$ is  $\lambda=\lambda_+ = |a|^2 /3$, when $w= v = -|a|/3$ and $u= (1+\sqrt{2})|a|/3$. Inserting this into the constraint 
\begin{equation}
k_1 u + k_2 w + k_3 v = 2N
\end{equation}
we find the corresponding value of the parameter $a$: 
\begin{equation}
a= a_+ = -\frac{6N}{ ( 1+\sqrt{2}) k_1 - k_2 - k_3}.
\end{equation}
Similarly, the lowest possible of $\lambda= \lambda_- = -|a|^2 /3$ implies $w= u = |a|/3$ and $v= -(1+ \sqrt{2})/3 $. The constraint on the solutions then implies that the corresponding value of parameter $a$ is 
\begin{equation}
a= a_- = -\frac{6N}{  k_1 + k_2 - ( 1+\sqrt{2}) k_3}.
\end{equation}

Since $ a_+ <0$, $k_1$ is restricted as 
\begin{equation}
k_1 > \frac{N\sqrt{2} }{1+\sqrt{2}}\approx \frac{N}{1.7071},
 \end{equation}
whereas, $a_- <0$ implies that  
 \begin{equation}
 k_3 < \frac{N\sqrt{2} }{1+\sqrt{2}}.
 \end{equation}
 
One therefore finds the following sets of possible degeneracies for the extrema with three distinct eigenvalues:
 1) For $N=2$, the only possibility is $(k_1, k_2, k_3) = (2,1,1)$, 2) For $N=3$ there are three solutions, with degeneracies $(k_1, k_2, k_3)$ equal to $(4,1,1) $, $(3,2,1)$, and $(2,3,1)$, 3) $N=4$: there are seven solutions, with $(k_1, k_2, k_3)$ equal to $(6,1,1)$, $(5,2,1)$, $(4,3,1)$, $(3,4,1)$, $(5,1,2)$, $(4,2,2)$, and $(3,3,2)$. For higher $N$ the number of possibilities increases. 

For each allowed set of degeneracies the corresponding solution exists in the interval $ a_- < a <a_+$ if $k_2 < 2N/(1+\sqrt{2})$, or otherwise in the interval  $a_+ < a < a_-$. At the boundary value $a=a_+$  (or at $a=a_-$) the three-distinct-eigenvalues extremum merges with the two-distinct-eigenvalues extremum $(k_1, k_2 + k_3)$ (or with $(k_1+k_2, k_3)$). At the same values of the parameter  $a=a_{\pm}$ the two-distinct-eigenvalue solutions change their character from a local minimum to a  saddle.

The large number of the extrema with three distinct eigenvalues notwithstanding, they are of less importance because they are all saddle points of the function $F(y)$.  It is enough to demonstrate this at one value of $a$ for which such an extremum exists, since the number of unstable directions around it can change only upon collision with other extrema, which happens only at the boundary values $a=a_{\pm}$. The simplest choice is then at an intermediate value of the tuning parameter $a$ for which the corresponding Lagrange multiplier is precisely $\lambda=0$. At this particular value the solution is $u=-v=2|a|/3$ and $w=0$. The function $F(y)$ near this point, call it $y_0$,  is 
\begin{equation}
F(y) = F (y_0) + \frac{1}{2} \delta y_i H_{ij} \delta y_j + O( \delta y ^3), 
\end{equation}
with the Hessian $H$ as a diagonal matrix with the $k_1 +k_3$ positive eigenvalues $2|a|$, and $k_2$ negative eigenvalues $-2 |a|$. Here $\delta y_i = y_i -y_{0,i}$, $i=1,2,...2N$. Let us assume first that $k_2 =1$, and chose $\delta y_{2N} = y_{2N} - w $. The constraint implies that 
\begin{equation}
\delta y_{2N} = -\sum_{i=1} ^{2N-1} \delta y_i,
\end{equation}
and therefore in the constrained subspace 
\begin{equation}
\frac{F(y) - F(y_0) }{|a|} = \sum_{i=1}^{2N-1} \delta y_i M_{ij} \delta y_j + O( \delta y ^3). 
\end{equation}
The $ (2N-1) \times (2N-1) $ stability matrix $M$ is given by
\begin{equation}
M= 1_{2N-1} - (2N-1) V \otimes V^T 
\end{equation}
with the normalized vector $V^T = (1,1,....1)/\sqrt{2N-1}$. Clearly $V$ is also an eigenvector of the stability matrix $M$, which therefore has the eigenvalues $2(1-N)$ (non-degenerate) and $1$ (degeneracy $ 2N-2 $). For $N\geq 2$ therefore $M$ has one unstable direction, and $y_0$ is a saddle. By continuity it then remains a saddle at all values of the parameter $a$ at which exists. 

If the multiplicity of the $w=0$ eigenvalue is $k_2 >1$, the unconstrained Hessian $H$ has less than  $2N-1$ stable directions, and therefore a  projection onto any $(2N-1)$-dimensional subspace cannot yield a local maximum. This proves that all extrema with three distinct eigenvalues are saddles, and as such can never be absolute minima of the function $F(y)$.   

\end{document}